\documentclass{article}

\usepackage{arxiv}
\usepackage[utf8]{inputenc} 
\usepackage[T1]{fontenc}    
\usepackage{hyperref}       
\usepackage{url}            
\usepackage{booktabs}       
\usepackage{amsfonts}       
\usepackage{nicefrac}       
\usepackage{microtype}      
\usepackage{graphicx}
\usepackage{acro}
\usepackage{array}
\usepackage{amsmath} 
\usepackage{authblk}

\usepackage{xcolor}
\usepackage{comment}
\usepackage{listings}
\usepackage{acro}
\usepackage{multirow}
\usepackage{pgf-umlsd}

\usepackage{orcidlink}
\newcolumntype{C}[1]{>{\centering\let\newline\\\arraybackslash\hspace{0pt}}m{#1}}
\newcolumntype{L}[1]{>{\raggedright\let\newline\\\arraybackslash\hspace{0pt}}m{#1}}

\DeclareAcronym{qkd}{
  short=QKD,
  long=Quantum Key Distribution,
}

\DeclareAcronym{qkdn}{
  short=QKDN,
  long=QKD Network,
}

\DeclareAcronym{kms}{
  short=KMS,
  long=Key Management System,
}

\DeclareAcronym{otp}{
  short=OTP,
  long=One-Time Pad,
}

\DeclareAcronym{pqc}{
  short=PQC,
  long=Post-Quantum Cryptography,
}

\DeclareAcronym{kma}{
  short=KMA,
  long=Key Management Agent,
}

\DeclareAcronym{ksa}{
  short=KSA,
  long=Key Supply Agent,
}

\DeclareAcronym{ckms}{
    short= CKMS,
    long= Carrier-KMS,
}

\DeclareAcronym{sdn}{
    short=SDN,
    long=Software-defined Networking,
}

\DeclareAcronym{gnmi}{
    short=gNMI,
    long=gRPC Network Management Interface,
}

\DeclareAcronym{akms}{
    short=AKMS,
    long= Access-KMS,
}

\DeclareAcronym{ukms}{
    short=UKMS,
    long= User-KMS,
}

\DeclareAcronym{qkdmod}{
    short=QKDM,
    long= QKD-Module,
}

\DeclareAcronym{aaa}{
    short=AAA,
    long=Authentication{,} Authorization and Accounting,
}

\DeclareAcronym{sae}{
    short=SAE,
    long=Secure Application Entity,
}

\DeclareAcronym{fcaps}{
    short=FCAPS,
    long={Fault, Configuration, Accounting, Performance and Security Management},
}

\DeclareAcronym{snmp}{
    short=SNMP,
    long=Simple Network Management Protocol,
}

\DeclareAcronym{euroqci}{
    short=EuroQCI,
    long=European Quantum Communication Infrastructure,
}

\DeclareAcronym{qbn}{
    short=CN,
    long=Carrier Network,
}

\DeclareAcronym{rng}{
    short=RNG,
    long=Random Number Generator,
}

\title{DemoQuanDT: A Carrier-Grade QKD Network}

\author[1, 6]{P. Horoschenkoff\orcidlink{0000-0001-9414-0765}}
\author[2]{J. Henrich\orcidlink{0009-0005-0915-9653}}
\author[3]{R. Böhn}
\author[4]{I. Khan\orcidlink{0000-0002-1376-2018}}
\author[1]{J. Rödiger\orcidlink{0000-0003-0832-8417}}
\author[3]{M. Gunkel\orcidlink{0000-0001-8853-1217}}
\author[2]{M. Bauch}
\author[4]{J. Benda}
\author[4]{P. Bläcker\orcidlink{0009-0008-5191-1898}}
\author[4]{E. Eichhammer}
\author[4]{U. Eismann\orcidlink{0009-0001-9971-2942}}
\author[3]{G. Frenck}
\author[5]{H. Griesser\orcidlink{0000-0002-2220-0172}}
\author[3]{W. Jontofsohn}
\author[1]{N. Kopshoff}
\author[1]{S. Röhrich}
\author[2]{F. Seidl}
\author[2]{N. Schark}
\author[4]{E. Sollner}
\author[3]{D. von Blanckenburg}
\author[2]{A. Heinemann\orcidlink{0000-0003-0240-399X}}
\author[2]{M. Stiemerling}
\author[3]{M. Gärtner}

\affil[1]{Rohde \& Schwarz Cybersecurity GmbH, Munich, Germany}
\affil[2]{Darmstadt University of Applied Sciences, Darmstadt, Germany}
\affil[3]{Deutsche Telekom Technik GmbH, Bonn, Germany}
\affil[4]{KEEQuant GmbH, Fuerth, Germany}
\affil[5]{Adva Network Security GmbH, Munich, Germany}
\affil[6]{Technische Universität München, Munich, Germany}

\begin{document}
\maketitle

\section*{Abstract}
Quantum Key Distribution Networks (QKDN) enable secure communication even in the age of powerful quantum computers. In the hands of a network operator, which can offer its service to many users, the economic viability of a QKDN increases significantly. The highly challenging operator-user relationship in a large-scale network setting demands additional requirements to ensure carrier-grade operation. Addressing this challenge, this work presents a carrier-grade QKDN architecture, which combines the functional QKDN architecture with the operational perspective of a network operator, ultimately enhancing the economic viability of QKDN. The focus is on the network and key management aspects of a QKDN while assuming state-of-the-art commercial QKD-Modules. The presented architecture was rolled out within an in-field demonstrator, connecting the cities of Berlin and Bonn over a link distance of 923~km across Germany. We could show, that the proposed network architecture is feasible, integrable, and scalable making it suitable for deployment in real-world networks. Overall, the presented carrier-grade QKDN architecture promises to serve as a blueprint for network operators providing QKD-based services to their customers.

\section{Introduction}\label{sec:introduction}

In modern communication systems, cryptography plays a vital role in establishing end-to-end secure channels, ensuring confidentiality, integrity, and authenticity of messages \cite{bsi_grundschutz_bausteine}. A cornerstone of this procedure is the secure derivation of a shared secret for encryption, which is typically accomplished through a two-step process. Asymmetric cryptographic algorithms, including RSA and ECC \cite{applied_crypto}, are initially employed to generate a key pair, typically by leveraging the public key infrastructure (PKI) for authentication between entities. The resulting asymmetric key pair serves as foundation for the subsequent secure negotiation of a symmetric key, which is then utilized to enable more resource-efficient encryption mechanisms. This hybrid paradigm is epitomized by the TLS protocol \cite{tls_protocol}. 

Classical, asymmetric cryptography relies on problems considered intractable for classical computing but faces significant vulnerability to advancements in quantum computation \cite{shor1994, Grover96, review_q_comp, bsi_entwickl_q_comp}, unlike symmetric cryptography and hashing, which are generally deemed more quantum-resilient \cite{bsi_krypto_verfahren}. The "harvest now, decrypt later" threat, where (quantum) computers could decrypt archived communications once sufficiently advanced, underscores the urgency for quantum-resistant cryptography \cite{nethen2024Pmmp}. Among others, this threat has recently led to guidelines advocating RSA deprecation by 2030 \cite{eu_pqc_recomm} and formal prohibition by 2035 \cite{nist_transition}, making migration a top priority \cite{bsi_pqc_recomm}.

Quantum-secure communication is mainly addressed via two principal strategies: \ac{pqc} and \ac{qkd}. \ac{pqc} employs algorithms based on computational problems resilient to both classical- and quantum attacks \cite{alagic_status_2022}, while \ac{qkd} leverages quantum mechanics and information theory to ensure physical-layer security \cite{amer2021qkd}. These approaches can be seen as complementary, with \ac{qkd} suited to high- and long-term security applications and \ac{pqc} offering practicality in broader, urgent transitions or more cost-sensitive scenarios \cite{qkd_pqc}.

As an optical technology, \ac{qkd} is prone to loss, facing inherent distance constraints. Thus, with optical fibers, the distance is usually limited to 100 km \cite{mehic2021qkdn}. Current satellite-based \ac{qkd} technologies extend this range but cannot fully replace fiber links due to other intrinsic limitations \cite{sat_comm_properties}. To address these challenges, quantum repeaters and trusted nodes are under active investigation, enabling secure end-to-end key establishment. Although quantum repeaters are constantly being advanced  \cite{q_repeat_review}, practical implementations of quantum repeaters are still an active research area, with on-going progress by the quantum optics community \cite{q_tele_internet}.

Consequently, prevailing implementations rely on trusted nodes for secure relay, where intermediate, point-to-point, keys enable end-to-end security. A network connecting multiple users by the \ac{qkd} technology, is referred to as a \ac{qkdn}. This raises the question of how to design, implement and maintain a \ac{qkdn} for ensuring key confidentiality, integrity and authenticity, cost efficiency, and reliability within the context of diverse application scenarios.

\subsection{Related Work}

Comparing the previously deployed \acp{qkdn} underlines the rapid improvement in this research and technology area \cite{pirandola2020advances, mehic2021qkdn, cao2022evolution}. The DARPA Quantum Network, established in 2002, was among the first to demonstrate the feasibility of \ac{qkd} in a network scenario, achieving a key rate of approximately 400~bit/s over 10 km leveraging 10 nodes \cite{darpa}. The 6 node SECOQC \ac{qkdn}, in Vienna, subsequently built on this foundation and besides increasing link distances and key rates, introduced a more complex multi-protocol infrastructure laying the groundwork for a layered trusted node network \cite{Peev_2009}. In 2010, the Tokyo \ac{qkdn} introduced a \ac{kms} into a hierarchical network management comprising six nodes in total \cite{qkdnTokyo}. Both the SECOQC and the Tokyo QKD Network were crucial in pushing forward standardization efforts needed to integrate \ac{qkd} into conventional telecommunication frameworks. 

In 2011, the SwissQuantum \ac{qkdn} showed advancements in terms of reliability by demonstrating a runtime of more than one-and-a-half years \cite{stucki2011long}. One of the largest \ac{qkdn} is notably the Beijing-Shanghai Quantum Communication Backbone, started to being build in 2014 \cite{qiu2014quantum}, reaching a length of 4,600~km \cite{Chen2021}. Additionally, the first satellite-based \ac{qkd} links were demonstrated as part of this network \cite{Liao2017, yin2017satellite}. In 2019, the Cambridge Quantum Network demonstrated its performance- and reliability capabilities of a high speed (2.58~Mb/s) and long-term (580 days) running \ac{qkdn} \cite{cambridge_qkdn}. 

A significant number of other \ac{qkdn} initiatives exist in the European Union (EU) \cite{padua_qkdn, kastrup2024interconnectedQkdn, poznan_qkdn, eagle_1_qkdn, madrid_nature, qu_net, euroqci2023}. Most notably, the European Quantum Communication Infrastructure (EuroQCI) initiative \cite{euroqci2023} and the QuNet Initiative in Germany \cite{qu_net}, focussing on the deployment of \ac{qkd} in real-world production networks between governmental institutions and other high-security environments in the EU and Germany. Among those, the Madrid Quantum Network \cite{madrid_early, madrid_nature}, has pushed for the adoption of the \ac{sdn} approach \cite{sdn_survey}, which is also widely used in classical networks. Later, this approach was transfered into an ETSI Standard \cite{etsi_015}.

However, current research has primarily focused on the use case where the network operator is identical to the network user, which is without loss of generality not the case in real-world scenarios. These networks, in which one operator is selling a provisioning service to multiple users, are referred to as "carrier-grade" networks.

In classical telecommunication networks, carrier-grade is synonymous with high reliability, availability, and performance in terms of the Quality-of-Service (QoS) concept. For \acp{qkdn}, adding the imperative of constant confidentiality, integrity and authenticity of key relays yields distinct requirements. The fusion of these requirements and the classical carrier-grade standards gives rise to a specific set of requirements for a carrier-grade \ac{qkdn}. The goal of this work is to lay the fundamentals for a carrier-grade \ac{qkdn} which could serve as blueprint for network operators aiming to achieve quantum secure communication by means of the current technology status. Throughout this work a secure channel is defined by its ability to ensure confidentiality, integrity and authenticitiy.

\subsection{Contributions}
In this paper, we design a carrier-grade \ac{qkdn} and start by defining core requirements, which we use as a foundation to devise and implement in a next step a concrete architecture. Security aspects, are addressed in a next step in which we identify architectural attack surfaces, propose corresponding countermeasures, and thereby extend an established security framework. Lastly, we realize this architecture in an in-field demonstrator that links the two German cities Berlin and Bonn over a distance of 923 km. We demonstrate the feasibility of integrating our solution within the carrier-grade context of a major European network provider, while meeting the defined requirements. Our findings provide a basis for us to further refine and enhance performance and security. In summary, the contributions of this paper are as follows:

\begin{enumerate}
    \item \textbf{Definition of requirements} for a secure and efficient archicture of a \ac{qkdn} in the context of a carrier-grade network infrastructure.
    
    \item \textbf{Design of a carrier-grade \ac{qkdn} reference architecture}, taking into account communication and interaction patterns for a comprehensive concept, based on identified requirements.

    \item \textbf{Comprehensive concept validation} based on a conceptual implementation, contingent upon the satisfaction of the defined requirements, the feasibility and integratability of the derived concept. 
\end{enumerate}

The paper is organized as follows: Section \ref{sec:requirements} details the requirements for a carrier-grade \ac{qkdn}, followed by the architectural design in Section \ref{sec:architecture}. Section \ref{sec:components} outlines key components, while Section \ref{sec:communication} elaborates communication patterns and security considerations. Section \ref{sec:experiment} evaluates the conceptual implementation, and Section \ref{sec:conclusion} concludes the work.

\section{Requirements}\label{sec:requirements}

Before outlining the requirements for developing a carrier-grade \ac{qkdn} architecture, we introduce the general concept of carrier-grade telecommunication networks and classify the two distinct services offered by \acp{qkdn}, providing essential context for understanding the proposed framework.

\subsection{Carrier-Grade Telecommunication Networks}\label{sec:cgtn}

In a carrier-grade telecommunication network, originally mentioned and defined around carrier-grade Ethernet for core network in the first decade of this century \cite{Kirstaedter},  aspects like performance, availability, operational efficiency and manageability as well as process conformity get into focus imposing stringent stipulations on operational carrier processes and carrier-hosted network elements. 

First, a carrier-grade telecommunication network relies on comprehensive monitoring and management tools to ensure easy operation and maintenance, providing visibility of all carrier-hosted network elements end-to-end. ITU-T recommendation M.3400 \cite{ITU3400} guides this effort and outlines five fundamental functionalities, namely \ac{fcaps} for effective network operation. Those functionalities are typically executed by a centralized network manager situated in a data center, interconnected with all network elements via a secure auxiliary classical Data Communication Network (DCN).

The employed network elements themselves must provide high data thoughput capacity at low latency and low power consumption. They must also show a high reliability with negligible downtime. In case of classical telecommunication networks, equipment capacity and availability expectations are in the range of multiple Tbit/s and less than a few minutes outage on average per year, respectively. The service experienced by the user shall be in the availability range of 99.999\% or "five nines". For a typical network this corresponds to an outage of less than 6 minutes per year \cite{five_nines}. 

Moreover, all network elements must match with established procurement and roll-out processes as typically applied in telecommunication providers. Amongst others, this comprises seamless interoperability across standardized interfaces, multi-vendor equipment sourcing, certificates of equipment conformity, compliance with central offices’ temperature models and physical dimensions of operated racks, as well as accessibility of well-trained and security-cleared technical staff for initial on-site and remote configuration and maintenance. 

At the networking level, automated redundancy mechanisms need to be implemented by backup resources. Protection or restoration switching capability must be in place to enable network recovery from arbitrary network failures. Furthermore, as one crucial carrier-grade ingredient, scalability refers to the capability of growing the network larger, i.e. connecting more and more nodes serving end users with growing traffic requests over the lifetime. And finally, another feature of a carrier-grade network is the connection to a higher-level Business Support Systems to be able to set up and manage customers and invoice them for the delivered service \cite{Kirstaedter}.

\subsection{Requirements for a Carrier-Grade QKDN}
\label{sec:req}

In view of the above described concept of a carrier-grade telecommunication network in general, we introduce requirements as a point of reference in the design and implementation process of a carrier-grade \ac{qkdn} architecture. The combination of common carrier-grade network requirements with the imperative necessity of persistent key confidentiality, integrity and authenticity gives rise to a specific set of requirements for a carrier-grade \ac{qkdn}. 

However, as existing operators' requirement lists for classical telecommunication networks might comprise more than a 1000 entries \cite{psa_classical_nets}, we restrict ourselves to a subset of requirements needed to enable the key provisioning service offered by \ac{qkdn} in a carrier-grade manner.

A \ac{qkdn} can operate in two distinct modes, which accommodate two different use cases. In the first use case, the user requests cryptographic keys and independently encrypts the data. In the second use case, the user transfers the data to the \ac{qkdn}, which then performs the encryption using the generated keys and organizes the data transfer. This work's focus is on the first use case, but to ensure mode flexibility, we have identified requirements that support both.

The requirements for a carrier-grade \ac{qkdn} were developed based on the state of the art in \ac{qkdn} specifications and architectures \cite{ITU-T-Y-3800, ITU3801, ITU3802, ITU3803, ITU3804, ITU3805}. These requirements, which were refined through discussions with a major European network operator, security and network equipment manufacturers, and universities, are summarized in Table \ref{tab:requirements}. The requirements are classified into four distinct categories. The term \textit{Functional} is used to describe requirements that are aimed at ensuring the basic provision of the service. The category \textit{Architecture} encompasses requirements that are designed to facilitate a meaningful conceptualization and implementation. The term \textit{Carrier-Grade} is used to describe requirements that originate from the context of traditional carrier networks and are intended to facilitate seamless integration and efficient operation. The category of \textit{Security} encompasses all aspects pertaining to the prevention of significant security related threats to the network or user data.

Within the scope of this work, a \ac{qkdn} satisfying the defined set of requirements is classified as a carrier-grade \ac{qkdn}. These requirements may provide a foundational reference for network operators aiming to achieve economically viable \ac{qkdn} deployment. In order to realize a fully-fledged carrier-grade \ac{qkdn}, the previously introduced requirements must be expanded. Supplementary requirements may be incorporated subsequently in future work to enhance functionalities or, for instance, to obtain third-party certification.

\begin{table}[htp]
    \centering
    \begin{tabular}{ |L{1.1cm}|C{3.5cm}|L{9.8cm}|} 
    \hline
     Abbrv. & Name & Description \\
     \hline
     \hline
     \multicolumn{3}{|c|}{\textbf{Functional}}\\
    \hline
     F\_KDE & Key Delivery for Encryption & The QKDN provides symmetric keys to remote users for their own encryption use. \\
     \hline
     F\_EEN & Efficient Encryption & The QKDN supports many users with high data rates (typically in the Gbit/s realm) at low cost \\
     \hline
     \hline
     
    \multicolumn{3}{|c|}{\textbf{Architecture}} \\
     \hline
     A\_NWN & Nation-wide \ac{qkdn} &  The QKDN overcomes reach limitations using trusted nodes to enable key sharing between non-adjacent or remote nodes. \\
     \hline
     A\_CSR & Compliant to Standards \& Recommendations & The QKDN follows established standards and recommendations for separating functions into distinct layers with clear interfaces. \\
     \hline
     \hline
    
    \multicolumn{3}{|c|}{\textbf{Carrier-Grade}} \\
    \hline
    C\_INT & Integration & The QKDN is integrated into the telecommunication providers' existing infrastructure. \\
    \hline
    C\_RBA & Role-based Access & The QKDN serves as a shared resource for multiple independent users, with operators and users having separate roles. \\    
    \hline
    C\_CCM & Centralized Controlling \& Management & The QKDN allows for a centralized control \& management, enabling a programmable, flexible and scalable network infrastructure. \\
    \hline
    C\_FCA & FCAPS & The QKDN includes FCAPS management with a focus on AAA, enabling secure and efficient service delivery to multiple end-users. \\
    \hline
    C\_MME & Modular Market Equipment  &  The QKDN avoids vendor lock-in, by using modular equipment with dedicated functionalities and interfaces, enabling interoperability among multiple vendors and supporting a multi-vendor environment. \\
    \hline
    \hline
    
    \multicolumn{3}{|c|}{\textbf{Security}} \\
    \hline
    S\_UIN & User Independent Network Security & The QKDN is secure against malicious users, as it does not receive or forward user-generated keys from outside the network. \\
    \hline
    S\_TRK & True Random Key & All derived keys are truly random, ensuring high entropy and unpredictability. \\
    \hline
    S\_NOP & Node Protection & A trusted node provides physical protection for stored assets, ensuring confidentiality, integrity, and availability against unauthorized access and environmental hazards.\\
    \hline
    S\_CRY & State-of-the-Art Cryptography & The QKDN uses state-of-the-art cryptographic protocols and primitives, integrated in a way that allows for crypto-agility to adapt to emerging threats. \\    
    \hline
    S\_COK & Confidentiality of the Key & The key is only accessible to communication devices and key forwarding components, requiring secure storage and authentication mechanisms in place. \\
    \hline
    S\_MUA & Mutual Authentication & Any communication involved in key forwarding process is mutually authenticated to prevent Man-in-the-middle (MiM)-attacks and to ensure authenticity.\\
    \hline
    \end{tabular}
    \caption{The requirements for a carrier-grade QKDN are grouped into "\textit{Functional}" for service provisioning, "\textit{Architecture}" for design and implementation, "\textit{Carrier-Grade}" for integration and operation within traditional carrier contexts, and "\textit{Security}" for mitigating network and data threats.}\label{tab:requirements}
    
\end{table}

\section{Architecture}
\label{sec:architecture}

To develop a carrier-grade \ac{qkdn} architecture that meets the defined requirements, we must consider additional aspects beyond the existing architectural fundamentals. We begin by reviewing the architectural principles outlined in the ITU-T recommendations, followed by describing the extensions we needed to make in fulfilling the previously defined requirements. A more detailed, communication and interaction among the various devices is covered in Chapter \ref{sec:components}.

\subsection{Functional Architecture}
In order to enable the integration of the \ac{qkd} technology into classical telecommunication networks different standards have been published \cite{ITU-T-Y-3800, ITU3801, ITU3802, ITU3803, ITU3804, ITU3805}, addressing among others the concepts, the functional requirements and -architectures, the key management as well as the \ac{sdn} paradigm of \ac{qkdn}. However, as a full review of all \ac{qkdn} published standards is out of scope for this work, this section selectively focuses on the indispensable ones crucial for comprehending the forthcoming concept. Remaining standard procedures and specified functions are, when needed, explained in the respective sections.

A \ac{qkdn} is a network comprised of two or more nodes connected through \ac{qkd} links \cite{ITU-T-Y-3800, ITU3801, ITU3802}. In accordance with requirement \textit{A\_NWN} we focus on technologies available today and omit the use of quantum repeaters. 

Trusted nodes are intermediate nodes, enabling the exchange of end-to-end keys between two or more users, which do not share a direct point-to-point connection. Distributing end-to-end keys, namely \ac{ksa} keys, to non-adjacent or remote users in the network can be visualized as a "horizontal" chain of trusted nodes with the users being on both ends. As in classical telecommunication networks, a \ac{qkdn} can be structured vertically into different hierarchical layers. The ITU-T Standards define the quantum layer, the key management layer, the application layer, the control layer, and the management layer \cite{ITU-T-Y-3800, ITU3801, ITU3802}. 

On the quantum layer, pairs of \acp{qkdmod} are connected via \ac{qkd} links, generating symmetric random bit strings according to a given \ac{qkd} protocol. The number of \acp{qkdmod} inherent in a single trusted node, usually scales with the number of adjacent trusted nodes, i.e. the node's degree. 

The layer above, known as the key management layer, is responsible for receiving and processing the bit string from the quantum layer in order to establish connections between the communicating parties. During processing, the received bit string is chopped up to generate keys. The keys resulting from this string are categorized as \ac{kma} keys. In general, these keys are necessary to provide the \ac{ksa} key and can be further refined as \ac{qkd} (generated) keys provided by the \ac{qkdmod} or other necessary keys provided by, e.g. a \ac{rng} situated in the \ac{kms}. The ITU-T defines a single \ac{kms} per trusted node, independent of the number of \acp{qkdmod}. \cite{ITU-T-Y-3800, ITU3801, ITU3802} 

Cryptographic applications using secure keys to encrypt and decrypt user traffic are located at the application layer. These cryptographic applications, in the ETSI Standard \cite{etsi_014} also termed \ac{sae}, request secure keys from the underlying key management layer. The application layer itself is not part of the \ac{qkdn}. It only interfaces with the \ac{qkdn} at its boundary.
The function of the management layer is to provide the network with \ac{fcaps} capabilities to support the management of the network of users. 
The control layer ensures proper operation of the \ac{qkdn}, by collecting the nodes' control status information and distributing actions. This layer includes routing algorithms to determine optimal paths for the key transport within the key management layer.

The ITU-T Standard Y.3803 \cite{ITU3803} deals with the functional elements of key management, the operations of key management and the key formats. Further, it provides an overview of current key distribution schemes. A key distribution scheme is realized by the \ac{kms}, which is responsible for delivering a secure key to ensure end-to-end encrypted data transmissions. This is achieved by forwarding a secure key through the trusted nodes. Hereby, four different schemes are distinguished. The schemes can be divided into two schemes using an additional centralized node and two schemes that forward the key without this centralized node. 

The ITU-T Standard Y.3804 \cite{ITU3804} specifies functions and procedures of the control and management of \ac{qkdn}. According to it, the \ac{qkdn} Controller oversees routing-, configuration-, policy-, access- and session control. The management of a \ac{qkdn} is performed identical to existent networks according to the \ac{fcaps} management. 

The ITU-T Standard Y.3805 \cite{ITU3805} provides a comprehensive framework for the integration of \ac{sdn} control paradigms into \ac{qkdn} architectures. This integration enhances network manageability and yields several key benefits, including centralized control for optimized monitoring and routing. Additionally, it enables dynamic programmability of tunable components, hierarchical scalability, and \ac{qkdn} virtualization, facilitating efficient resource sharing and rapid provisioning of secure communication services.

\subsection{Carrier-Grade Architecture}

\begin{figure}[htp]
    \centering
    \includegraphics[width=\linewidth]{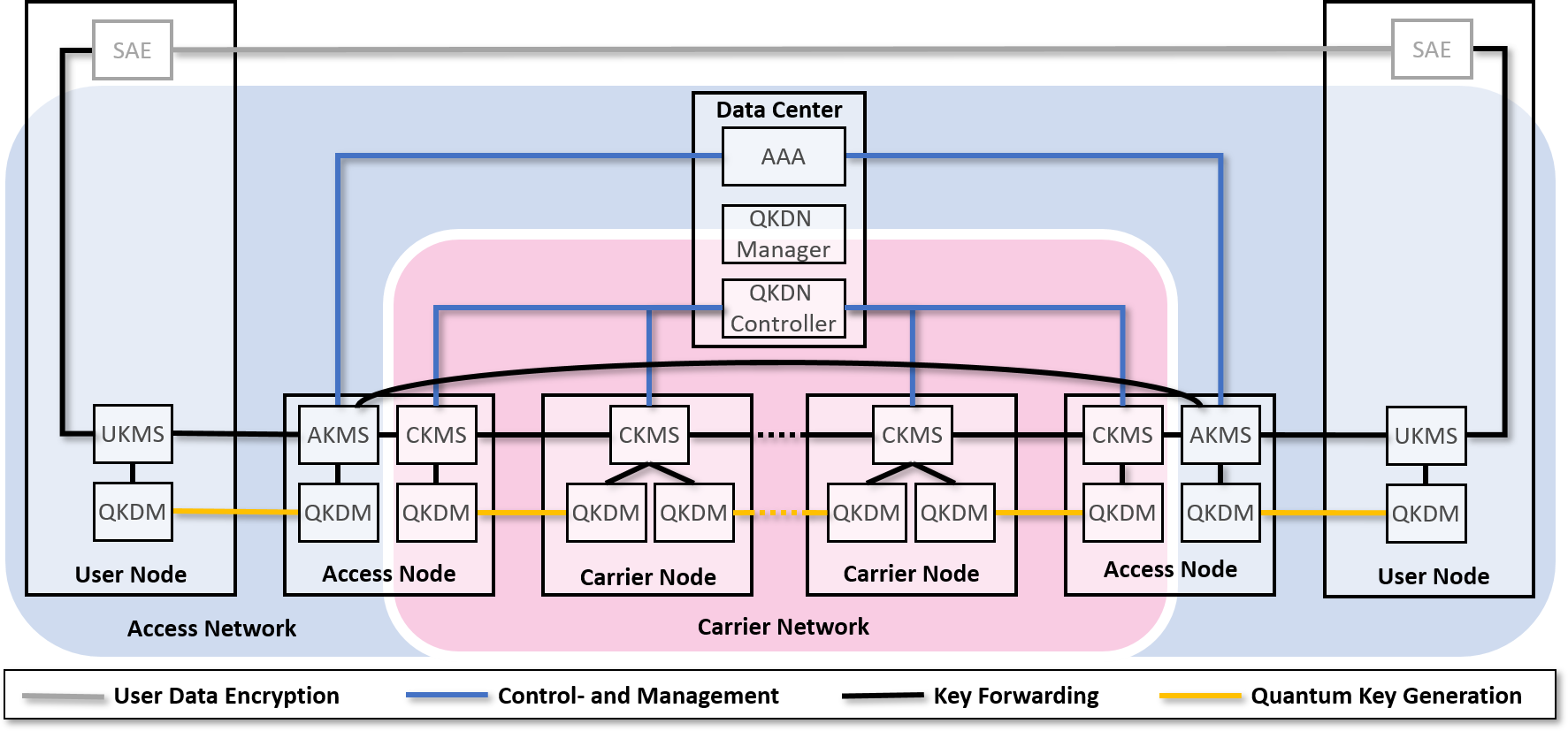}
    \caption{Schematic representation of the proposed architecture enabling a carrier-grade \ac{qkdn}. For meeting the requirements of a carrier-grade network, the single Trusted Node (TN) concept was split into three different TNs: the user node, the access node and the carrier node, all of them running a respective \ac{kms}, namely the user node a \ac{ukms}, the access node a \ac{akms} together with a \ac{ckms} and the carrier node a CKMS. This categorization enables the network operator to provide keys as a service. Beside the network controlling and managing functionalities realized by the QKDN Controller and QKDN Manager, the data center provides also the \ac{aaa} device. This network operator specific instance takes care of the user authentication, its authorization as well as accounting. Note, that as the QKDN Manager holds connections to any device entity in premise of the network operator, for the sake of clarity these are not depicted here. In this figure, the devices and nodes are grouped according to their processed information.}
    \label{fig:dq-architecture}
\end{figure}

A crucial characteristic of a carrier-grade \ac{qkdn} architecture is the role separation of the network operator and the network end-users, with the operator providing network services and the user consuming these. In accordance with requirement \textit{F\_KDE}, a carrier-grade \ac{qkdn} delivers a key for subsequent encryption executed by the user. As in existent carrier-grade networks, multiple users sharing the same network resources could exist, which might not necessarily trust each other, the user itself is not considered as part of the \ac{qkdn} (requirement \textit{S\_UIN}). Therefore, a carrier-grade \ac{qkdn} can be subdivided into two parts: an access network providing access to all individual users requesting a service, and a carrier (or backbone) network providing the fundamental key forwarding capacity between access nodes.

In order to integrate the \ac{qkdn} technology into the existent telecommunication provider's infrastructure and account for the different roles existent (requirement \textit{C\_INT} and \textit{C\_RBA}), we deploy nodes with varying privileges following the ITU-T -based approach (\cite{ITU3802}, Configuration 3). However, in accordance with the specified requirements in Table \ref{tab:requirements}, a precise delineation between access and carrier networks is established, accompanied by the allocation of specific functionalities to the distinct node types and other carrier-hosted elements.

The introduction of a user node, an access node, and a carrier node is contingent upon the specific functionality of the node in question. The user node resides at the end-users location and has minimum privileges. Higher privileges are attributed to access- and carrier nodes. The access nodes serve as intermediaries, bridging the access- and the carrier network. The carrier nodes, in turn, form the carrier infrastructure, enabling large-scale network topologies. The user node might be on the premise of a user, while the access- and carrier nodes are on premise of the operator.

\begin{figure*}[htp]
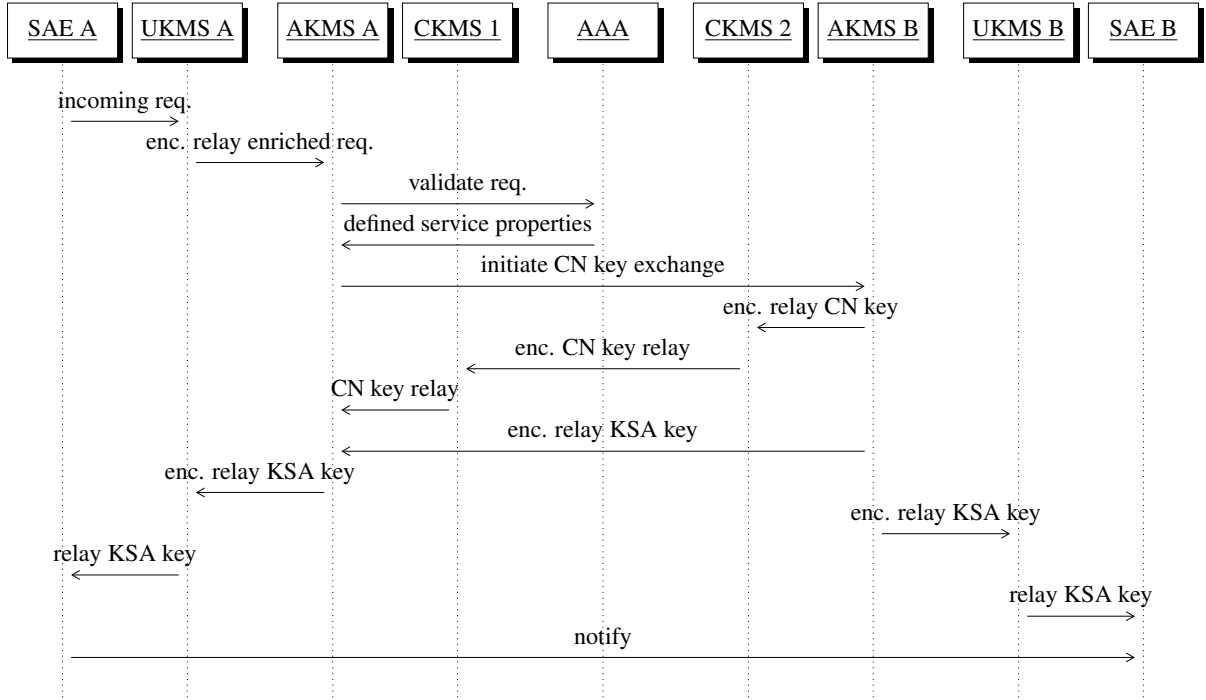

  \centering
   \resizebox{\linewidth}{!}{
  \begin{sequencediagram}
    \newinst{SAE_A}{SAE A}{}
    \newinst{UKMS_A}{UKMS A}{}
    \newinst[0.5]{AKMS_A}{AKMS A}{}
    \newinst{CKMS_1}{CKMS 1}{}
    \newinst[0.5]{AAA}{AAA}{}
    \newinst[0.5]{CKMS_2}{CKMS 2}{}
    \newinst{AKMS_B}{AKMS B}{}
    \newinst[0.5]{UKMS_B}{UKMS B}{}
    \newinst{SAE_B}{SAE B}{}

    \mess{SAE_A}{incoming req.}{UKMS_A}
    
    \mess{UKMS_A}{enc. relay enriched req.}{AKMS_A}
    
    \mess{AKMS_A}{validate req.}{AAA}
    
    \mess{AAA}{defined service properties}{AKMS_A}
    
    \mess{AKMS_A}{initiate CN key exchange}{AKMS_B}
    
    \mess{AKMS_B}{enc. relay CN key}{CKMS_2}
    
    \mess{CKMS_2}{enc. CN key relay}{CKMS_1}
    
    \mess{CKMS_1}{CN key relay}{AKMS_A}
    
    \mess{AKMS_B}{enc. relay KSA key}{AKMS_A}
    
    \mess{AKMS_A}{enc. relay KSA key}{UKMS_A}
    
    \mess{AKMS_B}{enc. relay KSA key}{UKMS_B}
    
    \mess{UKMS_A}{relay KSA key}{SAE_A}
    
    \mess{UKMS_B}{relay KSA key}{SAE_B}
    
    \mess{SAE_A}{notify}{SAE_B}

  \end{sequencediagram}
    }
  
  \caption{High-level schematic representation of key distribution process between two \acp{sae} within the carrier-grade \ac{qkdn} architecture, designed to accommodate both ETSI Standards 004 \cite{etsi_004} and 014 \cite{etsi_014} through a unified key ID and session management approach. The figure omits specific message exchanges, e.g. acknowledgements, error handling and QKDN Controller details. Abbreviations used: encrypted relays are referred to as "enc. relay" and requests as "req.".}
  \label{uml:kex}
\end{figure*}

According to the location and properties of the different node types, we categorize three different types of \ac{kms}: the \ac{ukms}, the \ac{akms} and the \ac{ckms}. The \ac{kms} types retain their fundamental characteristics as defined in \cite{ITU3803}, specifically the reception, processing, storing, and forwarding of \ac{qkd} keys, but exhibit distinct variations in their additional functionalities. To describe these functionalities, we outline in the following the basic principle of the key distribution process, which involves the complete \ac{kms} chain, c.f. Fig. \ref{fig:dq-architecture}. A detailed description outlining their roles and interfaces is given in a reference-like framework in Section \ref{sec:components}. It should be noted that, in this architecture, newly introduced cryptographic keys are named according to their usage rather than their generation origin. For sake of clarity, this process is depicted in a simplified UML diagramm in Fig. \ref{uml:kex}.

A key request is initiated by an \ac{sae} in a user node, hereby referred to as \ac{sae} A. Consequently, the request is received by the \ac{ukms} A, which then directs it, enriched with user related information, to its \ac{akms}. This \ac{akms} is from now on referred to as \ac{akms} A, c.f. Fig. \ref{uml:kex}. The \ac{akms} A forwards the request to the \ac{aaa} device, which is a carrier-specific device for request validation and described later. By comparing the request with the contractually defined service properties a validation is performed. A prominent example of the validation process would be the verification of whether valid payment information is available. Upon successful validation, the \ac{akms} A contacts the corresponding \ac{akms} in the receiving access node, referred to as \ac{akms} B, c.f. Fig. \ref{uml:kex}. Hereby, information on the key request, as number of keys, key size and involved communication parties is exchanged allowing for defined storage initialization and management routines.

The information from the requesting \ac{sae} B, to all necessary communication parties involes a translation between the \acp{sae}, the \ac{ukms} devices and the \ac{akms} devices. This translation is facilitated by the prior exchange with the \ac{aaa} device. A process which can be compared to using a telephone directory, referencing the \ac{sae} via the corresponding \ac{ukms} to the respective \ac{akms}.

The contacted, \ac{akms} B now generates a \ac{kma} key, according to requirement \textit{S\_TRK}. This generated \ac{kma} key is in this work referred to as "\textit{\ac{qbn} key}". Subsequently, the \ac{qbn} key is, together with the receiving \ac{ckms} ID information, forwarded to the first \ac{ckms}, which relays the key back to the \ac{akms} A in the known decentralized hop-by-hop manner through the \ac{ckms} chain, as shown in Figure 6 in \cite{ITU3803}. In accordance with requirement \textit{S\_NOP}, cryptographic measures maintaining confidentiality are not prerequisites for the communication link between the \ac{akms} and the \ac{ckms}.

By deploying the hop-by-hop key distribution scheme, we are able to fulfill the requirements \textit{S\_CRY} and \textit{F\_EEN}, as this scheme might show more flexibility in terms of cryptoagility by supporting both schemes, the One-Time-Pad as well as the AES-GCM cipher \cite{aes_gcm_tls}.

With the secure connection in place, leveraging the \ac{qbn} key, the \ac{akms} B creates the \ac{ksa} key as per requirement \textit{S\_TRK}, and transmits it securely over this channel. The secure \ac{akms}-\ac{akms} communication channel is illustrated by the bow-shaped black line connecting the two \ac{akms} in Fig. \ref{fig:dq-architecture}.

In the final step, the \ac{ksa} key is securely transferred from the \ac{akms} devices to the respective \ac{ukms} devices, which in turn forward the key to the \acp{sae} for secure encryption. Throughout the key distribution protocol, stringent security controls and procedures are enforced to guarantee the secure management of cryptographic keys, fulfilling the \textit{S\_COK} and \textit{S\_CRY} requirements.  

The network's control and management layer is enabled by connecting each node type to a central data center according to requirement \textit{C\_CCM}. The data center hosts three devices: the \ac{qkdn} Controller, the \ac{qkdn} Manager and the \ac{aaa} device. The task of the \ac{qkdn} Controller is to control all \ac{qkdn} entities hosted in the carrier network, e.g. the \acp{qkdmod} and the \ac{ckms} devices, as forseen in \cite{ITU3804, ITU3805}. This includes the optimal path selection a key takes in the carrier network. The \ac{qkdn} Manager ensures network management tasks in terms of the \textit{C\_FCA} requirement \cite{ITU3400}. As part of the \ac{fcaps} functionalities, the authentication, authorization and accounting of users is performed by the \ac{aaa} device. 

The introduced network architecture comes with the following security advantages. Firstly, since the \ac{akms} serves as a demarcation device between the access network and the carrier network, efficiently seperating these two domains, as it does not allow for metadata gathering on the \ac{akms} for the carrier network, as this information is by-passed and not accessible to the users. In particular, the \ac{akms} safeguards the carrier network from the unauthorized access by users hiding metadata, including the carrier network’s topology, from the user and keeping the user information isolated from the carrier network. This reduces the overall network’s vulnerability in the event that a user node is compromised, as it is not part of the carrier network. Secondly, the key distribution process, only allows keys that are passed on exclusively from the core of the \ac{qkdn} to the users and not vice versa.

The design of a symmetric network architecture facilitates rapid scalability and incorporates the inherent symmetry of the \ac{qkd} technology, whereby identical cryptographic keys are established between two remote parties concurrently. As we built upon the ITU-T architectural concept, we fulfilled the \textit{A\_CSR} requirement. Lastly, it is important to note that by defining only the interfaces, i.e. the interaction of the devices, we allow a multi-vendor strategy, as is common in classic telecommunication networks (requirement \textit{C\_MME}).

The network setup phase is typically managed by the network operator and the device providers, and is therefore outside the scope of this work. It is likely that initially the requirement \textit{S\_MUA} is achieved during this phase through specific means, e.g. by the use of smartcards anyway needed for activating the various certified devices. 

Summarized, for supporting the derived requirements a carrier-grade \ac{qkdn} shall fulfill, we introduced three different \ac{kms} types and established a novel key distribution scheme.

\section{Components}
\label{sec:components}
This section contains a detailed explanation of the introduced components, highlighting their respective tasks and interfaces and outlining their roles and responsibilities within the carrier-grade \ac{qkdn} architecture. Its purpose is to serve as a quick and sorted reference framework to detail the components introduced in Section \ref{sec:architecture}. An overview of the different interfaces of the introduced \ac{kms} is given in Table \ref{tab:matrix_recommend}. Importantly, each \ac{kms} operates as a separate entity with its own unique security requirements, as detailed in Section \ref{sec:communication}.

\subsection{User-KMS}

The \ac{ukms} is situated within each user node and establishes connections with at least one \ac{sae}, the \ac{akms}, and the \ac{qkdmod} as described in Table \ref{tab:matrix_recommend}. Notably, each \ac{sae} is connected to a single \ac{ukms}, whereas a \ac{ukms} may maintain multiple connections with distinct \ac{sae}. The task of the \ac{ukms} is to receive and process user requests, communicate with the respective \ac{akms}. In detail, the following process, triggered by an incoming user request, is carried out: the received requests are forwarded to the connected \ac{akms}. The request formulation process yields an enriched request, combining the information stipulated in each standard with necessary user information for effective \ac{aaa} device validation. 

Messages received from the \ac{akms} contain the requested \ac{ksa} key. Depending on the network configuration, keys may be buffered in advance. The communication between the \ac{ukms} and the \ac{akms} is secured by the keys that are generated between the \acp{qkdmod} in the respective nodes. Additional \ac{kms} related operations beyond those mentioned, like storage routines and management interfaces, are governed by the ITU-T Standard \cite{ITU3803}.

\subsection{Access-KMS}

The \ac{akms} is deployed within the access node, where it interfaces with the local \ac{ckms}, with the \ac{ukms} entities located in the neighboring user nodes, \ac{akms} entities in other access nodes, the \ac{aaa} device, the \ac{qkdn} Manager, and the \ac{qkdmod} connected to the user nodes, c.f. Table \ref{tab:matrix_recommend}. As primary interface to the carrier network, the \ac{akms} acts as a network gateway, separating the carrier network from the user network. Its operational scope encompasses the validation of \ac{ukms}-initiated key requests and the forwarding of key requests to the attached \ac{ckms}. Further, it may also incorporate synchronization of key storages with peer \ac{akms} entities.

The key distribution process in the carrier network is triggered by the reception of a key request from the \ac{ukms}, and entails the following sequence of events: after successfully validating the received request with the defined service properties provided by the connected \ac{aaa} device, the \ac{akms} A contacts \ac{akms} B, c.f. Fig. \ref{uml:kex}. Hereby, information about the key request is exchanged. This includes the number and size of requested keys, IDs of the involved \acp{sae}, the \ac{ukms} B and sending \ac{ckms} IDs. In contrast to the previous communication with the \ac{aaa} device, no defined service paramters are exchanged.

The \ac{akms} B triggers the key distribution protocol in the colocated \ac{ckms}, as defined in the ITU-T \cite{ITU3802}. Specifically, the \ac{akms} B first generates the \ac{qbn} key and passes it, together with the sending \ac{ckms} ID, to the \ac{ckms}. The \ac{qbn} key can be exchanged unencrypted, as both are colocated. Upon receiving the \ac{qbn} key from the colocated \ac{ckms} in the \ac{akms} A, an acknowledgement is returned. 

In a next step, the \ac{akms} B creates a \ac{ksa} key and transmits it securely to the \ac{akms} A, utilizing the previously exchanged \ac{qbn} key for encryption. In a last step, both \ac{akms} push the \ac{ksa} key towards the \ac{ukms}. Further \ac{kms} related operations are performed, as defined in the ITU-T Standard \cite{ITU3803}.

The generation of the \ac{ksa} key and the \ac{qbn} key in the \ac{akms} comes mainly with two advantages. Firstly, both \ac{akms} already have an established channel for initiating the request communication. An exchange of the \ac{ksa} key between both \ac{ckms} would leverage the need for another channel. Secondly, the \ac{akms} already needs a \ac{rng} source for creating the \ac{qbn} key, which can be reused here. However, in specific cases, outsourcing either or both key generation tasks from the \ac{akms} to the \ac{ckms} may prove beneficial, e.g. for balancing the security load between the two components.

\subsection{Carrier-KMS}
The carrier network is build from the entirety of \ac{ckms}, which provide long-distance key relaying. The \ac{ckms} devices are distributed across access and carrier nodes, with their functional profiles being independent of their situated node type, functioning as a "\ac{kma} key exchange platform". 

The \ac{ckms} devices located in carrier nodes are connected to at least one other \ac{ckms} in a carrier-node, the \ac{qkdn} Controller, the \ac{qkdn} Manager, the respective \acp{qkdmod} and the \ac{akms}. The \ac{ckms} devices residing in the access nodes are additionally connected to the \ac{akms}, see also Table \ref{tab:matrix_recommend}. 

The operational scope is defined by the functions and procedures specified in the ITU-T Standard \cite{ITU3803}. The previously generated \ac{qbn} key is forwarded in a hop-by-hop manner along an optimal path through the carrier network.

\subsection{QKDN Controller, QKDN Manager and AAA}

The QKDN Controller, QKDN Manager, and \ac{aaa} device are co-located within the data center, which is interconnected with the \acp{akms}, \acp{ckms}, and \acp{qkdmod}. The \ac{qkdn} Controller is responsible for executing several key functions, including: facilitating interoperability between devices operating at the key management and application layer to enable seamless communication. The separation between the AAA device and the \ac{qkdn} Controller is in line with the separation between access- and carrier network.

In the context of developing an \ac{sdn}-capable \ac{qkdn} according to \cite{ITU3805}, the \ac{qkdn} Controller continuously receives monitoring data from node-internal devices, courtesy of the \ac{sdn}-Agent present in every node. This monitoring data encompasses information relevant to the routing process, such as the number of available keys, the rate of key generation, and contractually defined parameters that are incorporated in the path optimization. Utilizing this knowledge, the \ac{qkdn} Controller performs optimal routing based on criteria, such as minimizing latency, maximizing throughput, or ensuring security policies. For a reactive routing protocol this can be triggered by the \ac{ckms} residing in the access node. In addition to routing capabilities, the \ac{qkdn} Controller has additional orchestration tasks, including network automation and network configuration. 

The \ac{qkdn} Manager is tasked with realizing the \ac{fcaps} functionalities, including authenticating existing network instances, monitoring their behavior, and interacting with the \ac{qkdn} Controller in the event of device failure, as well as conducting alarm management. \cite{ITU3400, ITU3804}

As part of the \ac{fcaps} functionalities, the \ac{aaa} device is responsible for implementing the \ac{aaa} functionalities on the user-layer, which are essential for providing services to users. It performs a validation of the user information forwarded by the \ac{akms}, verifying its correctness and legitimacy. Moreover, it manages the translation tables that enable interoperability between the \ac{sae}, the \ac{ukms} and \ac{akms}, thereby facilitating the key distribution process.


\begin{table*}[htp]
    \centering
    \begin{tabular}{| L{1cm} || L{1.8cm} | L{7.5cm} | L{3.5cm} |}
        \hline

         Device & Interface & Task & Implementation \\
        \hline
        \hline

         \multirow{4}{*}{\ac{ukms}} & \ac{sae} & Receive key requests from one or multiple users &  ETSI GS QKD 014 \cite{etsi_014} \\

          & \ac{akms} & Forward user information and request related information; KSA key relay &  RFC 9113, HTTP/2 \cite{http2_protocol}   \\
          
          &  QKD Module & Ensure encryption between the UKMS and the AKMS &  ETSI GS QKD 014 \cite{etsi_014} \\
        \hline
        \hline

        \multirow{4}{*}{AKMS} & AKMS & Initiate \ac{qbn} key exchange & RFC 9113, HTTP/2 \cite{http2_protocol} \\
        
         & \ac{aaa} & Validate user requests; reception of defined service properties & RFC 1157, SNMP \cite{snmp_protocol}  \\
         
          & QKDN Manager & Authentication and monitoring of device & RFC 1157, SNMP \cite{snmp_protocol}\\
        \hline
        \hline
        
         \multirow{5}{*}{CKMS} & \ac{akms} & \ac{qbn} key exchange between carrier and access networks & RFC 9113, HTTP/2 \cite{http2_protocol} \\
         
          & CKMS & \ac{qbn} key exchange in carrier network & RFC 9113, HTTP/2 \cite{http2_protocol} \\
          
          & QKDN Controller & Determine next optimal CKMS hop; Exchange of status information & RFC 9113, HTTP/2 \cite{http2_protocol} and gNMI \cite{gnmi} \\
          
          & QKDN Manager & Authentication and monitoring of device &  RFC 1157, SNMP \cite{snmp_protocol} \\
        \hline
         
    \end{tabular}
    \caption{KMS interfaces resulting from the carrier-grade \ac{qkdn} architecture presented in Sec. \ref{sec:architecture}. In order to enable a secure data exchange, supplementary measures and protocols are needed, c.f. Section \ref{sec:communication}. Further, \ac{akms}-\ac{ukms} and \ac{ckms}-\ac{akms} interfaces are not included due to the presence of redundancy. Description of \ac{akms} or CKMS to \ac{qkdmod} is also omitted, as these correspond to the interface between \ac{ukms} and \ac{qkdmod}.}

    \label{tab:matrix_recommend}
\end{table*}

\section{Security Aspects} \label{sec:communication}

This section is concerned with the assurance of secure communication between the various components. While a detailed security target would require, among others, a well-defined attacker model, which is beyond the scope of this work, this section aligns with and builds upon existing efforts with similar objectives, pointing out important attack surfaces that need to be considered in securing the proposed carrier-grade \ac{qkdn} architecture.

In this context, the ITU-T X.1710 Standard \cite{ITU-T-X-1710} provides a foundational security framework for the layered structure of a \ac{qkdn} as defined by ITU-T. As discussed in Section \ref{sec:architecture}, the proposed architecture adheres to the ITU-T Standards. Building upon the security objectives and security threats in the ITU-T X.1710 Standard, we extend this framework by addressing the handling of user data beyond the application layer and emphasizing architectural peculiarities such as the prevention of unauthorized user access to the carrier network.

\subsection{Information Assets}

The exchanged data in the proposed architecture is classified into four distinct categories, according to their distinct security requirements. The initial categorization is based on ITU-T Standard X.1710 \cite{ITU-T-X-1710}.

\emph{Key data} comprises cryptographic material requiring high confidentiality, integrity and authenticity to maintain the \ac{qkdn}'s core functionality. This data is exchanged between the \acp{qkdmod}, all \ac{kms} instances, and \acp{sae}, with heightened sensitivity for connections within trusted nodes. 

\emph{Meta data} includes key management information necessary for efficient key establishment. It requires confidentiality, integrity, and authenticity and is exchanged among the same entities as key data. Excluding this information from the \ac{qkdn} Controller is beneficial upon relaying on non-centralized key distribution schemes.

\emph{Control and Management Information} pertains to network management, including routing and performance data. It demands at minimum integrity, and authenticity to prevent network manipulation, with its volume and quality depending on the routing protocol. This information is exchanged between the \ac{ckms} and the \ac{qkdn} Controller, as well as between the components attached to the \ac{qkdn} Manager.

In addition to the three categories already defined, we introduce a fourth category: \emph{network user profile data}. Network user profile data is managed within the \ac{kms} layer in the carrier-grade \ac{qkdn} architecture, which clarifies the necessity of this as an additional category and does not fall into the previously defined categories. This category encompasses information related to the carrier network’s customers and their communication needs. This data should be treated with a high level of security, ensuring confidentiality of the networks users and their contractual parameters. Communication in this category occurs between the \ac{akms} and the \ac{ukms}, the \ac{akms} and the \ac{akms}, as well as the \ac{akms} and \ac{aaa} device. This further expands to possible back-end mechanisms and infrastructure systems, e.g. for node recovery.

\subsection{Security Measures}

The selection of the security measures depends on the strength of the assumed attacker, particularly the extent to which they can actively or passively interfere with the process, and the computational resources at their disposal. 

This architectural concept emphasizes two critical peculiarities: the role of the \ac{akms} and \ac{aaa} device in preventing unauthorized user access to the carrier network and their specific inter-\ac{kms} interfaces. The \ac{akms}, serving as the primary gateway to the user node, is co-located with the \ac{ckms} within the same physical node. To ensure high-security properties akin to a firewall, physical separation is to be implemented by using distinct hardware solutions. Additional device hardening measures include closing unnecessary ports and services, utilizing security audit tools, and adopting other established security mechanisms. Given its role in accessing network user profile data, the \ac{aaa} device necessitates robust security measures similar to those implemented in the \ac{akms}. To prevent vulnerabilities, the communication interfaces between these two devices must be rigorously defined and secured, eliminating any potential for unguarded access points. 

The specific inter-\ac{kms} interfaces govern communication between the following instances: \ac{akms}-\ac{aaa} and \ac{akms}-\ac{akms}. In line with ongoing standardization efforts on an inter-\ac{kms} interface, only minimal data should be visible to adversaries on encrypted channels. Upon initiation of the network setup phase, a process typically governed by the network operator's proprietary protocols and aligned with the network equipment manufacturer's technical specifications, the initial secure communication channel between \ac{akms} entities is established. This secure channel may be facilitated through the utilization of a pre-shared key, e.g. by secure exchange via smart-card technology during the provisioning phase of the devices. Subsequently, this initial communication framework can be seamlessly transitioned to use the key provisioning service of \ac{qkdn} as part of an additional security extension protocol. The communication interface between the \ac{aaa} device and the \ac{akms} involves control and management traffic, whose security provisions fall outside the purview of this present work. Methods for ensuring security at this layer can be found in \cite{security_cm}. Nevertheless, it is imperative that both interfaces adhere to the customary security requirements that are typically mandated for \ac{kms} channels, as delineated in the established guidelines of the ITU-T Standard X.1710. 

In alignment with a clear separation between the access and carrier network, a connection from the \ac{akms} to the \ac{qkdn} Controller is to be omitted due to network and security separation concerns. This demarcation, between the access and carrier network, should be preserved in any case, including future architectural improvements.  

Secure communication within this framework is contingent upon the presence of a reliable \ac{rng}, which facilitates the generation of both \ac{qbn} and \ac{ksa} keys. There are different approaches to categorize \acp{rng} and their functional and security properties. One established approach is \cite{bsi_rng} which categorizes into distinct functionality classes, each requiring the availability of statistical models and tests as per device specifications to attain specific security classes. Notably, PTG.3 represents the most robust functionality class, employing a hybrid approach that combines a physical entropy source (compliant with PTG.2) with computational post-processing to ensure a higher level of security. This hybrid design provides a fail-safe mechanism, guaranteeing at least computational security in the event of physical source failure  \cite{bsi_rng}. To ensure secure network operation, the deployment of a PTG.3-compliant \ac{rng} is mandatory. This category can be realized using a Quantum Random Number Generator (QRNG) as an entropy source, followed by post-processing techniques to ensure the fail-safe mechanism. Additionally, this generation process is based on a stochastic model, which is a necessary basis for the PTG.3 classification.

In this section, ITU-T-defined security assets were extended with an additional fourth category, aligned with the carrier-grade \ac{qkdn} architecture. A thorough examination identified four main architectural attack surfaces, which need to be considered for achieving a highly secure \ac{qkdn} and an inquiry into targeted countermeasures against adversaries was undertaken.

\section{Implementation}\label{sec:experiment}

\begin{figure}[htp]
    \centering
    \includegraphics[width=\linewidth]{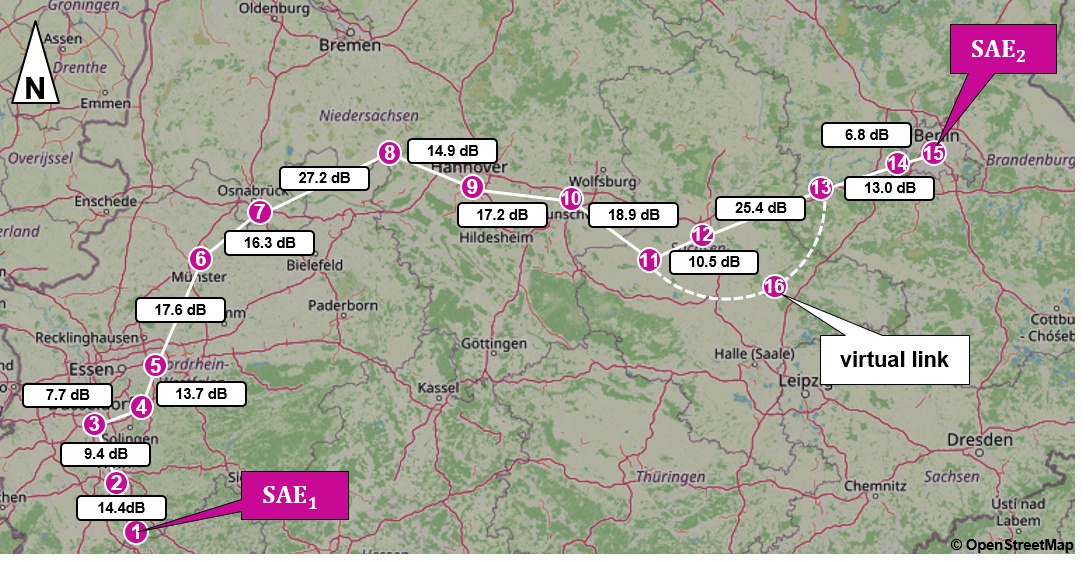}
    \caption{Layout of the DemoQuanDT Network. It facilitates a large-scale \ac{qkdn} between the cities Berlin and Bonn. The trusted node based \ac{qkdn} consists of 15 nodes bridging a distance of 923 km. The network follows the previous described carrier-grade \ac{qkdn} architecture. A virtual link (node 16) can be enabled for testing the fault management procedure in case of a \ac{qkd} link failure. Moreover, the carrier-grade service may be utilized by multiple \ac{sae} at both ends of the network, thereby showcasing the network's ability to support a diverse range of applications.}
    \label{fig:dq_map}
\end{figure}

The network architectural concept detailed in the preceding sections was successfully demonstrated through an in-field demonstrator in Germany, establishing a 923 km quantum-secured connection between Berlin and Bonn \cite{DemoQuanDT}. To the best of our knowledge, this is the first carrier-grade \ac{qkdn} and the largest network of its kind in Germany.

In order to demonstrate the feasibility and applicability of the derived concept and being able to evaluate its quality, we implemented a \ac{qkdn} between the cities of Berlin and Bonn which incorporates the principles of the introduced carrier-grade \ac{qkdn} architecture. The physical layout of this in-field demonstrator, developed in close collaboration with a major European network operator, is depicted in Fig. \ref{fig:dq_map}.

The primary focus of this in-field demonstrator is on the novel key distribution process within the proposed architecture, enabling an in-depth examination of system interoperability, as detailed in Section \ref{sec:components}. The two sites at the respective end points constitute the user nodes, each accommodating a dedicated \ac{sae} realized by network encryptors. The network's key provisioning service may be utilized by multiple \ac{sae} at both ends of the network.

The access nodes are positioned at the node preceding each of the two endpoints, thereby facilitating the interconnection of the user nodes with the carrier network. The remaining nodes collectively form the carrier network infrastructure, providing the necessary connectivity and routing capabilities. \ac{kms} devices are deployed according to the respective node types. Overall the network consists of 16 nodes, whereby one node is entirely virtual enabling route switching. The deployed standards and interfaces of the introduced components are listed in Table \ref{tab:matrix_recommend}. 

Integrability was achieved by incorporating the design roll-out within the established standards of the collaborating network operator. This mainly involved provisioning both the node and user sites, the corresponding fiber infrastructure, as well as hosting the data center including the respective interfaces to the network elements. Network manufactures, network operators and universities delivered and developed the components essential for a carrier-grade \ac{qkdn}, c.f. \ref{sec:components}. 

This included new opportunities for standardization of an inter-\ac{kms} interface. Defining this interface enables the use of \ac{kms} from multiple vendors, supporting the network operator's multi-vendor strategy. The conceptual implementation has successfully put into practice the defined \ac{kms} interface, with multiple vendors successfully integrating and interacting their \ac{kms} systems, showcasing seamless multi-vendor interoperability. 

The demonstrator's control and management layer was intentionally designed with a focused scope, allowing for a concentrated emphasis on key aspects of the carrier-grade \ac{qkdn} architecture. This resulted in a conceptual network controller as well as in the simplifying assumption that all user requests are legitimate and authenticated, thereby rendering the explicit validation process of the \ac{aaa} device superfluous. The integration of a \ac{qkdn} into the standard operations of a network operator aligns with established roll-out processes for experiments and demonstrators, including internal network management protocols. In this conceptual setup, the \ac{akms} and \ac{ckms} are deployed on the same hardware, relying on software-based separation to differentiate between the carrier and access network functionalities. 

At each node, off-the-shelf \acp{qkdmod} from multiple vendors, or simulated \ac{qkdmod} were deployed, c.f. Table \ref{tab:dq_devices}. The simulated \ac{qkdmod} enabled deterministic key provisioning at a set key rate, ensuring, that the quantum-layer extends across all nodes. This allowed acquisition of meaningful data at the upper \ac{qkdn} layers and especially in the key management layer, our primary goal.

As the network constitutes a virtual path, it is possible to perform routing on a small scale. Laying the fundamentals for future QoS mechanisms, the routing is executed by a link-weight-customized protocol triggered by the \ac{ckms} located in the access node. Upon request, the optimal path is calculated by the developed \ac{qkdn} Controller, who is sequentially updating the nodes' routing tables along the optimal path. Following the \ac{sdn} approach, the optimal path is determined by periodic updates from the connected \ac{ckms} instances, showcasing a clear separation of the control and data plane. After the initiating \ac{ckms} receives the successfull path generation acknowledgement, the hop-by-hop relay is executed.

A first, exemplary measurement of the constructed network assesses the time required to exchange a key between the two \acp{sae}, i.e. $T_{key}$. The average $T_{key}$ value is determined from over 10,000 measurements, assuming a fully populated key storages, i.e. no delay originating from the quantum layer. Within the conceptual implementation, the measured $T_{key}$ of around $17.34 \pm 1.62$ seconds, supports an encrypted user traffic of around 180 Gbit/s, leveraging the AES-256 cipher at the highest packet-to-key ratio of 389 GB per 256-bit key \cite{aes_crypto_cipher}. Notably, the current $T_{key}$ is a result of the initial demonstrator's focus on feasibility and functionality, rather than high-speed optimization, leaving room for future improvements in this area.

Summarized, our in-field demonstrator suggests that the proposed network concept is not only theoretically feasible but also practically integrable into the existing roll-out processes of a network operator. This significant finding demonstrates the potential acceptance and scalability of the architecture and underlines the potential for smooth and efficient use in real network environments, proving its practicality.

\begin{table}
    \centering
    \begin{tabular}{| c | c | c | c | c |}
        \hline
         Node & KM Layer & Link & SKR [kb/s] & QBER [\%] \\
        \hline
        \hline
        1   & UKMS &  \multirow{15}{*}{\begin{tabular}{c}
            1-2\\
            2-3\\
            3-4\\
            4-5\\
            5-6\\
            6-7\\
            7-8\\
            8-9\\
            9-10\\
            10-11\\
            11-12\\
            12-13\\
            13-14\\
            14-15
            \end{tabular}}

        & \multirow{15}{*}{\begin{tabular}{c}
            - \\
            - \\
            - \\
            $2.0\pm0.1$ \\
            - \\
            - \\
            $22.7\pm3.7$\\ 
            $1.2\pm0.1$\\ 
            $0.3\pm0.1$\\ 
            $21.3\pm1.9$\\ 
            $2.2\pm0.1$\\ 
            $11.8\pm2.6$\\ 
            $2.0\pm0.4$\\
            -
        \end{tabular}} 
        &  \multirow{15}{*}{\begin{tabular}{c}
            -\\
            -\\
            -\\
            $2.4\pm0.5$\\
            - \\
            - \\
            $5.4\pm0.3$\\
            $1.7\pm0.5$\\
            $1.6\pm0.5$\\
            $4.5\pm0.3$\\
            $1.6\pm0.5$\\
            $5.6\pm0.4$\\
            $1.3\pm0.6$\\
            -
            \end{tabular}} \\
        2   & A/CKMS &   &   & \\
        3   & CKMS &   &   &  \\
        4   & CKMS &   &   &  \\
        5   & CKMS &   &   & \\
        6   & CKMS &   &   & \\
        7   & CKMS &   &   & \\
        8   & CKMS &   &   & \\
        9   & CKMS &   &   & \\
        10  & CKMS &   &   & \\
        11  & CKMS &   &   & \\
        12  & CKMS &   &   & \\
        13  & CKMS &   &   & \\
        14  & A/CKMS &   &  & \\
        15  & UKMS &   &  & \\

        \hline
    \end{tabular}
    \caption{Average Secret-Key-Rate (SKR) and average Quantum-Bit-Error-Ratio (QBER) values are calculated from measurement data collected over a one-week period at 30s intervals. For link 1-2, key data could be polled from the \ac{qkdmod}, but no SKR or QBER data was captured. For the remaining links, currently, simulated \acp{qkdmod} are deployed.}
    \label{tab:dq_devices}
\end{table}

\section{Conclusion} \label{sec:conclusion}
Enabling the key provisioning service on a large scale, commensurate with the requirements of a network operator, is essential for the economic viability of \ac{qkdn}. This work presents a novel carrier-grade \ac{qkdn} architecture that synergistically integrates the \ac{qkdn} architecture with a network operator service perspective. This integration was accomplished by mapping the concept of a carrier-grade network to the unique key provisioning service of a \ac{qkdn}, thereby ensuring a high level of key security, service reliability, while maintaining scalability. The resulting list of requirements is divided into four main categories. Functional for service provisioning, architecture for design and implementation, carrier-grade for integration and operation and security for mitigating network threats.

In order to fulfill these requirements a carrier-grade \ac{qkdn} architecture is developed, further extending existing standards. The proposed concept is characterized by the introduction of three \ac{kms} types, each situated at its corresponding node type, namely the \acf{ukms}, \acf{akms}, and \acf{ckms}. The \ac{ukms}, located at the user node, receives and processes incoming requests, which are subsequently forwarded to the \ac{akms}. The \ac{akms} functions as a demarcation point between the access and carrier networks, effectively obscuring the user's view of the network topology. The \ac{ckms}, situated within the carrier network, relays the key in a hop-by-hop manner, ensuring secure key distribution. 

A salient feature of this segragation is that the \ac{ksa} key, required for encrypting user traffic, is exclusively handled within the access network, thereby ensuring that keys can only be passed from the core of the \ac{qkdn} to the users and not vice versa. The network architectural concept outlined in the preceding sections was subjected to an in-field prototype \ac{qkdn} conducted across Germany, wherein a 923 km quantum-secured connection was successfully established between the cities of Berlin and Bonn. This network is to our best knowledge this is the first carrier-grade \ac{qkdn} and the largest network of its kind in Germany. To validate the feasibility of the proposed concept, we conducted an initial experiment within the DemoQuanDT network infrastructure. The results confirmed the practicality of the derived architectural concept and paved the way for further experiments, including investigations into routing protocols.

\newpage

\section*{Funding} 
The research stems from the German research ministry "Bundesministerium fuer Bildung, Wissenschaft, Forschung und Technologie" (BMBF) as part of the DemoQuanDT research and innovation project under grand agreement No. 16KISQ074.

\section*{Acknowledgments} 
We would like to thank Marc Fischlin and Sebastian Clermont from TU Darmstadt for the useful discussions and comments that have significantly improved this manuscript.

\section*{Competing Interests}
The authors declare no competing interests.

\section*{Data Availability Statement} 
Data underlying the results presented in this paper can be made available upon reasonable request.

\section*{Correspondence and Requests}
Correspondence and requests should be given to the consortium leader Marcus Gärtner [marcus.gaertner@telekom.de].

\end{document}